\documentstyle[12pt,epsfig,amssymb]{article}
\begin{document}
\setlength{\parskip}{0.45cm}
\setlength{\baselineskip}{0.75cm}
\begin{titlepage}
\begin{flushright}
CERN-TH/98-56 \\ 
DTP/98/06 \\
February 1998 \\
\end{flushright}
\vspace{0.1cm}
\begin{center}
\Large
{\bf {Polarized $\boldmath{\Lambda}$}-Baryon Production in} 

\vspace{0.1cm}
{\bf $\boldmath{pp}$ Collisions} 

\vspace{1.2cm}
\large
D.\ de Florian$^a$, M.\ Stratmann$^b$, W.\ Vogelsang$^a$\\

\vspace*{1.5cm}
\normalsize
{\it $^a$Theoretical Physics Division, CERN, CH-1211 Geneva 23, Switzerland}

\vspace*{0.1cm}
{\it $^b$Department of Physics, University of Durham, Durham DH1 3LE, England}

\vspace{1.5cm}
%
\large
{\bf Abstract} \\
\end{center}
\vspace{0.5cm}
We study the production of longitudinally polarized $\Lambda$-baryons
in single-spin $p\vec{p}$ collisions at RHIC and HERA-$\vec{N}$ as a means of 
determining the spin-dependent $\Lambda$ fragmentation functions. 
It is shown that a measurement of the rapidity distribution of the 
$\Lambda$'s would provide an excellent way of clearly discriminating between 
various recently suggested sets of polarized $\Lambda$ fragmentation 
functions that are all compatible with present $e^+e^-$ data. 
We also address the main theoretical uncertainties, which appear to
be well under control.
\end{titlepage}
\newpage
%
%
\noindent
The understanding of spin-dependent deep-inelastic scattering processes
(DIS) in terms of QCD--evolved polarized parton distributions
$\Delta f (x,Q^2)$ ($f=q,\bar{q},g$) is still far from being 
satisfactory, despite significant experimental and theoretical 
progress over the past few years. In particular the angular momentum 
component of the proton's spin and the polarized gluon density 
$\Delta g(x,Q^2)$ remain almost completely unknown for the time being, 
and more experimental results are required. 

Studies of spin-transfer reactions could provide further invaluable and
completely new insight into the field of ``spin physics'' and, in addition,
might also yield a better understanding of the hadronization process.
Such cross sections can be expressed as convolutions of perturbatively
calculable partonic spin-transfer cross sections with certain sets of 
parton distributions and fragmentation functions, whose scale dependence 
is completely predicted by QCD once a suitable non-perturbative input 
at some reference scale has been determined by data. To obtain a 
non-vanishing twist-2 spin-transfer asymmetry, the measurement of 
the polarization of one outgoing particle is obviously required, in addition 
to having a polarized beam or target. This certainly provides a great 
experimental challenge. $\Lambda$-baryons are particularly suited for
such studies due to the self-analyzing properties of their dominant weak
decay $\Lambda \rightarrow p \pi^-$, and recent results on $\Lambda$
production reported from LEP~\cite{data} have demonstrated the experimental 
feasibility of successfully reconstructing the $\Lambda$ spin.

In \cite{dsv} a first attempt was made to determine the spin-dependent 
$\Lambda$ fragmentation functions by analyzing these LEP data~\cite{data} 
in leading and next-to-leading order QCD, using the results of a preceding
study of unpolarized $\Lambda$ fragmentation functions.
Unfortunately it turned out, however, that the available LEP data,
all obtained on the $Z$ resonance and hence only sensitive to the 
flavor non-singlet part of the cross section, cannot even sufficiently
constrain the valence fragmentation functions for all flavors. Rather
different, but all physically conceivable, scenarios adopted for the
input valence distributions appear to describe the data equally well, and
for the ``unfavoured'' sea quark and gluon fragmentation functions one has to 
fully rely on mere assumptions. Clearly, further measurements of other
helicity transfer processes are required to test the models proposed 
in \cite{dsv}.

With the advent of RHIC \cite{rhic}, spin transfer reactions can be 
studied for the first time also in $pp$ scattering at c.m.s.\ energies of 
up to $\sqrt{s}=500\,\mathrm{GeV}$. In the following we will demonstrate that
such measurements would provide a particularly clean way of 
discriminating between the various conceivable sets of 
spin-dependent $\Lambda$ fragmentation functions presented in \cite{dsv},
and are almost unaffected by theoretical uncertainties. 
For this purpose, only {\em one} polarized
beam at RHIC would be needed. It should be noted here that similar 
(and almost equally useful) measurements could be performed also in a 
possibly forthcoming experiment at DESY, HERA-$\vec{N}$ \cite{heran}, 
utilizing the existing polarized ``fixed'' gas target of HERMES and the 
{\em{un}}polarized HERA proton beam.

%
The process we are interested in is $p\vec{p}\rightarrow \vec{\Lambda} X$,
the arrows denoting a longitudinally polarized particle. For the
time being, the required partonic helicity transfer cross sections, i.e.,
$q\vec{q}\rightarrow q \vec{q}$, $\ldots$, $g\vec{g}\rightarrow g \vec{g}$,
are calculated only to leading order (LO) accuracy and can be found, 
for example, in 
\cite{gw}. Hence we have to restrict our analysis to LO, implying the use of
LO-evolved $\Lambda$ fragmentation functions, contrary to the case of 
$e^+ \vec{e}^- \rightarrow \vec{\Lambda} X$ or SIDIS ($e \vec{p}\rightarrow 
\vec{\Lambda} X$, $\vec{e} p \rightarrow \vec{\Lambda} X$) where all relevant
coefficient functions  are now available at next-to-leading order (NLO) 
\cite{singlepol,doublepol,thesis,dsv}. In various analyses of processes 
sensitive to polarized parton distributions it has turned out to be 
particularly useful to study distributions differential in the rapidity
of a produced particle \cite{rapidity}, to which we therefore limit 
ourselves also in the present analysis.

The relevant differential polarized cross section can be schematically 
written as (the subscripts ``$+$'',``$-$'' below denote 
helicities)
\begin{eqnarray}
\label{eq:cross}
\frac{d\Delta \sigma^{p\vec{p}\rightarrow \vec{\Lambda} X}}{d \eta}
&\equiv& \frac{d\sigma^{pp_+\rightarrow \Lambda_+ X}}{d \eta} -
\frac{d\sigma^{pp_-\rightarrow \Lambda_+ X}}{d \eta} \\
&&\hspace*{-2.5cm} = \int_{p_T^{min}} \hspace{-0.2cm} dp_T 
\sum_{ff'\rightarrow i X} 
\int dx_1 dx_2 dz\, f^p(x_1,\mu^2) \times \Delta f'^p(x_2,\mu^2) \times
\Delta D_i^{\Lambda}(z,\mu^2) \times 
\frac{d\Delta \sigma^{f\vec{f'}\rightarrow \vec{i}X}}{d\eta} \; ,
\nonumber
\end{eqnarray}
the sum running over all possible LO subprocesses, and where we have 
integrated over the transverse momentum $p_T$ of the $\Lambda$, with 
$p_T^{min}$ denoting some suitable lower cut-off to be specified below.
The $\Delta f^p$ ($f^p$) are the usual (un)polarized parton distributions
of the proton, and 
\begin{equation}
\label{eq:ffdef}
\Delta D_i^{\Lambda}(z,\mu^2) \equiv D_{i(+)}^{\Lambda (+)}(z,\mu^2) -
 D_{i(+)}^{\Lambda (-)}(z,\mu^2)
\end{equation}
describe the fragmentation of a longitudinally polarized parton $i$ into
a longitudinally polarized $\Lambda$, where $D_{i(+)}^{\Lambda (+)}(z,\mu^2)$ 
$(D_{i(+)}^{\Lambda (-)}(z,\mu^2))$ is the probability for finding a $\Lambda$ 
at a mass scale $\mu$ with positive (negative) helicity in a parton 
$i$ with positive helicity, carrying a fraction $z$ of the parent parton's
momentum.

The directly observable quantity will be not the cross section in 
(\ref{eq:cross}) itself but the corresponding spin asymmetry,
defined as usual by
\begin{equation}
\label{eq:asym}
A^{\Lambda} \equiv \frac{d \Delta \sigma^{p\vec{p}\rightarrow \vec{\Lambda}X}/
d \eta}{d\sigma^{pp\rightarrow \Lambda X}/d \eta}
\end{equation}
where the unpolarized cross section 
$d\sigma^{pp\rightarrow \Lambda X}/d \eta$ is given by an 
expression similar to the one in (\ref{eq:cross}), with all $\Delta$'s
removed.

To study the sensitivity of (\ref{eq:asym}) to the poorly known 
$\Lambda$ fragmentation functions $\Delta D_i^{\Lambda}$, we use the three
LO sets obtained in \cite{dsv}. For the discussion below the idea
behind these very different models for spin-dependent 
$\Lambda$ fragmentation should be briefly recalled here:
 
\noindent
{\bf{Scenario 1}} is based on expectations from the non-relativistic 
naive quark model, where only strange quarks can contribute to the 
fragmentation processes that eventually yield a polarized $\Lambda$.

\noindent
{\bf{Scenario 2}} is inspired by estimates of Burkardt and 
Jaffe \cite{burkjaf,jaffe2} for a fictitious DIS structure function 
$g_1^{\Lambda}$, taking into account a similar breaking of the 
Gourdin-Ellis-Jaffe sum rule \cite{gej} for $\Lambda$'s as is observed
for nucleons. Assuming the same features also for the 
$\Delta D_i^{\Lambda}$, a sizeable negative contribution from $u$ and $d$ 
quarks to $\Lambda$ fragmentation is predicted here.

\noindent
{\bf{Scenario 3}} is the most extreme counterpart of scenario 1 since all
the polarized fragmentation functions are assumed to be equal here, which 
might be realistic if, for instance, a sizeable contribution to the 
production of polarized $\Lambda$'s results from decays of heavier hyperons 
who have inherited the polarization of $u$ and $d$ quarks produced originally.

For the unpolarized parton distributions of the proton, $f^p$, appearing in 
(\ref{eq:cross}), (\ref{eq:asym}) we use the LO set of Ref.\ \cite{grv} 
throughout our calculations (using other recent LO sets would not lead 
to any sizeable differences here). 
Unless otherwise stated we use for the corresponding polarized densities
$\Delta f^p$ the LO GRSV ``standard'' scenario \cite{grsv}.
For the unpolarized 
$\Lambda$ fragmentation functions $D_i^{\Lambda}$ needed for calculating  
$d\sigma^{pp\rightarrow \Lambda X}/d \eta$ we use the LO set 
presented in \cite{dsv}, which provides an excellent description of all
available, rather precise $e^+e^-$ data. It should be emphasized, however, 
that there are still sizeable uncertainties for the $D_i^{\Lambda}$,
mainly related to possible $SU(3)_f$ breaking effects not discernible 
from the presently available data. We note that in contrast to this,
the assumption of $SU(2)_f$ symmetry ($D_u^{\Lambda}=D_d^{\Lambda}$) appears 
to have a far more solid foundation. Clearly, further measurements of the
$D_i^{\Lambda}$ are required here. Nevertheless, the uncertainty in the 
$D_i^{\Lambda}$ resulting from $SU(3)_f$ breaking does not really affect 
our conclusions to be drawn below, since the contribution from strange quark 
fragmentation to the unpolarized cross section is only about 5$\%$.

Fig.~1(a) shows our predictions for the spin asymmetry $A^{\Lambda}$ as a
function of rapidity, calculated according to Eqs.\ (\ref{eq:asym}) and
(\ref{eq:cross}) for $\sqrt{s}=500\,\mathrm{GeV}$ and $p_T^{min}=13\,
\mathrm{GeV}$. Note that we have counted positive rapidity in the 
forward region of the {\em polarized} proton. We have used the three 
different scenarios for the $\Delta D_i^{\Lambda}$ discussed above, 
employing the hard scale $\mu=p_T$. The ``error bars'' 
should give an impression of the achievable statistical accuracy for 
such a measurement at RHIC. They have been estimated via
\begin{equation}
\label{eq:err}
\delta A^\Lambda \simeq \frac{1}{P} \frac{1}
{\sqrt{b_{\Lambda} \epsilon_{\Lambda} {\cal{L}}\,
\sigma^{pp \rightarrow \Lambda X}}} \;\;\; ,
\end{equation}
assuming a polarization $P$ of the proton beam of about 70$\%$,
a branching ratio $b_{\Lambda}\equiv BR(\Lambda\rightarrow p \pi)\simeq 0.64$,
a conservative value for the $\Lambda$ detection efficiency of 
$\epsilon_{\Lambda}=0.1$, and an integrated luminosity of 
${\cal{L}}=800\,\mathrm{pb}^{-1}$ \cite{rhic}. The cross section
$\sigma^{pp \rightarrow \Lambda X}$ is the unpolarized one, integrated over 
suitable bins of $\eta$. It should be mentioned that results almost 
identical to the ones in Fig.\ 1(a) can be obtained also for a
lower c.m.s.\ energy of $\sqrt{s}=200\,\mathrm{GeV}$ and a correspondingly
lowered $p_T^{min}$ and luminosity of 8 GeV and $240\,\mathrm{pb}^{-1}$,
respectively. Fig.~2(a) shows our results for a conceivable future
measurement at HERA-$\vec{N}$ at a much lower energy 
$\sqrt{s}=40\,\mathrm{GeV}$ and for $p_T^{min}=4\,\mathrm{GeV}$ 
and ${\cal{L}}=240\,\mathrm{pb}^{-1}$ \cite{heran}. It should be stressed 
that the $p_T$ cuts we have introduced do not only guarantee
the applicability of perturbative QCD (the hard scale $\mu$ in 
(\ref{eq:cross}) should be ${\cal{O}}(p_T)$), but also ensure 
that finite-mass corrections to the cross section, which would become
increasingly important for $z\leq 0.05$, remain small \cite{dsv}.
Furthermore, small values of $z$ also have to be excluded in order to 
make sure that there are no unreasonably large NLO contributions: as was 
noticed for the DIS case in \cite{dsv}, the (unpolarized) NLO kernels
for the evolution of the fragmentation functions have an extremely singular 
behaviour at small $z$, which eventually must lead to a complete breakdown 
of the ``perturbative'' formalism we use.

The behaviour of $A^{\Lambda}$ in Figs.\ 1(a) and 2(a) for the different
sets of polarized $\Lambda$ fragmentation functions can be easily understood 
from the fact that the process, in this particular kinematical region, is 
dominated by contributions from $u$ and $d$ quarks, so that the 
differences between the predictions in Figs.\ 1(a) and 2(a) are driven by the 
differences in the corresponding $\Delta D_u^{\Lambda}$ and
$\Delta D_d^{\Lambda}$. This immediately implies that the asymmetry 
has to be close to zero for scenario 1, negative for scenario 2 and 
positive and larger for scenario 3. The $\eta$-dependence is also 
readily understood: at negative $\eta$, the parton densities of the 
polarized proton are probed at small values of $x_2$ (i.e., in the 
``sea region''), where the ratio $\Delta q(x_2)/q (x_2)$ is also 
small. On the contrary, at large positive $\eta$, typical values of 
$x_2$ correspond to the valence region where the quarks are polarized much 
more strongly, resulting in an asymmetry that increases with 
$\eta$.

The results in Figs.\ 1(a) and 2(a) clearly demonstrate the usefulness of such
kind of measurements to determine the polarized $\Lambda$ fragmentation 
functions more precisely. The expected statistical errors are much smaller 
than the differences in $A^{\Lambda}$ induced by the various models. 
Thus an analysis of $A^{\Lambda}$ would provide an excellent way of 
ruling out some of the presently allowed sets of spin-dependent $\Lambda$ 
fragmentation functions, {\em provided} the observed differences in 
$A^{\Lambda}$ are not obscured or washed out by the theoretical 
uncertainties inherent in this calculation. We will therefore finally address 
this important point in some detail to demonstrate that the uncertainties 
appear to be well under control for this particular process and do not
impose any severe limitations.

There are three major sources of uncertainties: the dependence of $A^{\Lambda}$
on variations of the hard scale $\mu$ in (\ref{eq:cross}), which is of
particular importance since we are limited to a LO calculation, our present 
inaccurate knowledge of the precise $x$-shape and the flavor
decomposition of the polarized densities $\Delta f^p$, 
especially of $\Delta g$, and our ignorance of 
$\Delta D_g^{\Lambda}$ which is not constrained at all by the 
presently available $e^+e^-$ data \cite{dsv}. Fig.~1(b) gives an example of 
the scale dependence of $A^{\Lambda}$ by changing the scale from $\mu=p_T$
to $\mu=p_T/2$ for scenario 3. The same is shown in Fig.~2(b) for the 
HERA-$\vec{N}$ situation and scenario 2. Even though $d\Delta\sigma/d\eta$ 
and $d\sigma/d\eta$ individually change by as much as a factor 2 at certain
values of $\eta$, the uncertainty almost 
cancels in the ratio $A^{\Lambda}$. This gives us some confidence that the 
(unknown) NLO corrections might also cancel to some extent in the asymmetry, 
a pattern observed for all available NLO corrections involving polarized 
particles. We also show in the same figures the changes in the predictions 
resulting from varying the polarized parton distributions, using the recent 
LO set 1 of Ref.\ \cite{dss}, denoted by DSS, instead of the GRSV 
\cite{grsv} one. As can be observed, the asymmetry remains practically
unchanged, and differences can only be found at the end of 
phase space (at large values of $\eta$) where the cross section becomes 
small anyway. Also, as an extreme way of estimating the impact of the  
polarized gluon distribution, we have artificially set it to 
zero $(\Delta g (x,\mu^2) \equiv 0)$. We find
that changes in our predictions only occur in the region of negative 
$\eta$, but are small in the interesting region $\eta>0$ where the 
asymmetries are larger.
 
Finally, in order to examine the role played by $\Delta D_g^{\Lambda}$ in our 
analysis, we have used two different approaches: the standard one for
our polarized fragmentation functions, where the polarized gluon fragmentation
function is assumed to be vanishing at the initial scale \cite{dsv} 
and is then built up by evolution (``std. $\Delta D_g^{\Lambda}$''), and a set 
corresponding to assuming $\Delta D_g^{\Lambda} \equiv D_g^{\Lambda}$ at 
the same initial scale of Ref.\ \cite{dsv} (``max. $\Delta D_g^{\Lambda}$'') 
while keeping the input quark fragmentation functions unchanged. 
As can be observed, the resulting differences are also negligible, again 
due to the fact that $u$ and $d$ fragmentation dominate.

\section*{Acknowledgments}
The work of one of us (DdF) was partially supported by the World Laboratory.
\newpage
%

%
%
\section*{Figure Captions}
\begin{description}
\item[Fig.\ 1] {\bf (a)} The asymmetry $A^{\Lambda}$ as defined in 
(\ref{eq:asym}) as a function of rapidity of the $\Lambda$ at RHIC energies
for the various sets of spin-dependent fragmentation functions.
The error bars have been calculated according to (\ref{eq:err}) and
as discussed in the text.
{\bf (b)} same as for scenario 3 in {\bf (a)}, but using the ``maximal'' 
$\Delta D_g^{\Lambda}$ (see text), a hard scale $\mu=p_T/2$, 
$\Delta g=0$, or the 
spin-dependent parton distributions of the proton of set 1 of \cite{dss}. For 
comparison the solid line repeats the original result for scenario 3 of 
{\bf (a)}.
\item[Fig.\ 2] {\bf (a)} Same as Fig.~1(a), but for HERA-$\vec{N}$ kinematics.
{\bf (b)} Same as Fig.~1(b), but for HERA-$\vec{N}$ kinematics and scenario 2.
\end{description}
\newpage
%
\pagestyle{empty}
\begin{center}

\vspace*{-1.0cm}
\hspace*{-1.cm}
\epsfig{file=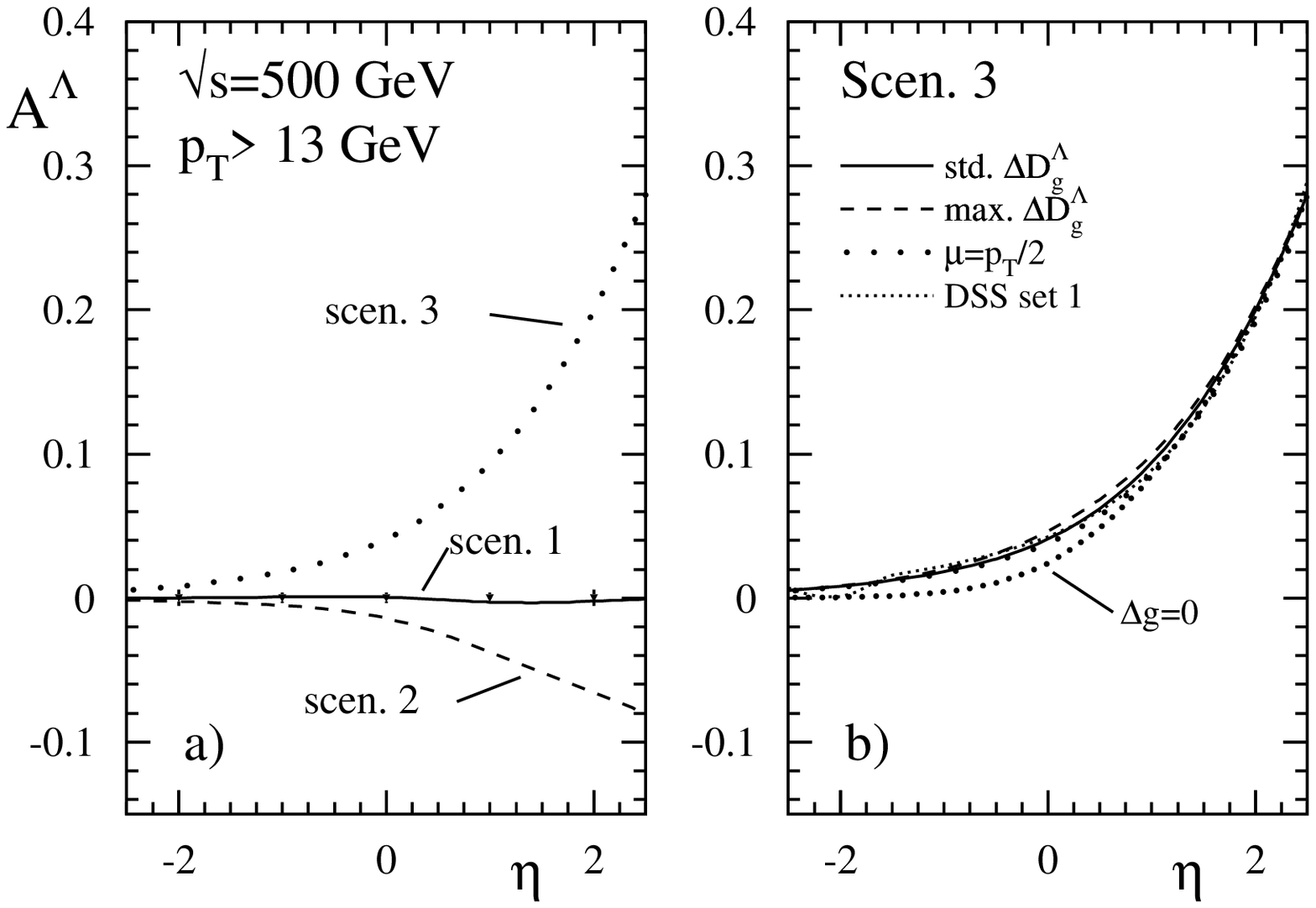}

\vspace*{-1.2cm}
\Large{\bf{Fig.\ 1}}
\end{center}

\begin{center}

\vspace*{-1.0cm}
\hspace*{-1.cm}
\epsfig{file=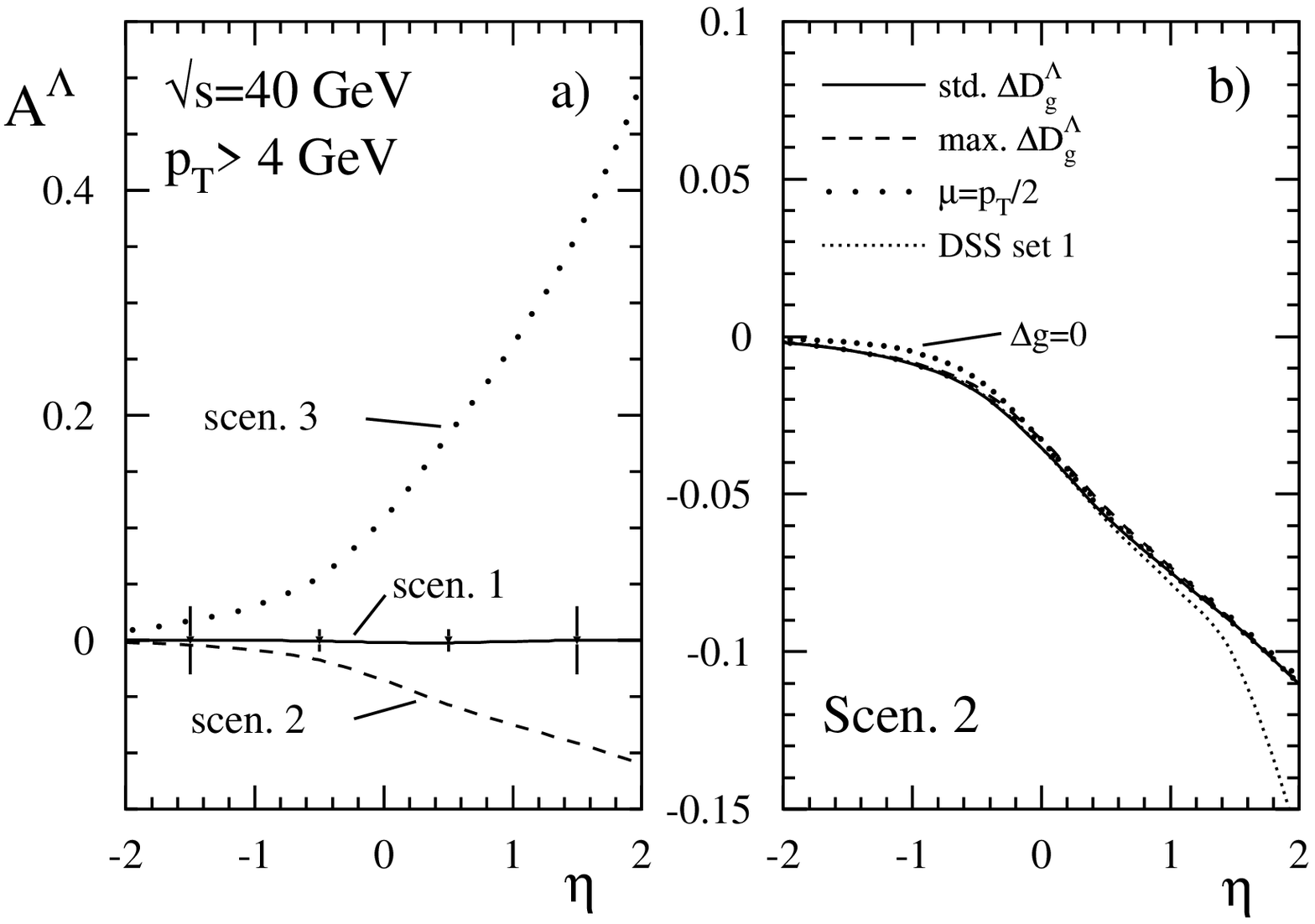}

\vspace*{-1.2cm}
\Large{\bf{Fig.\ 2}}
\end{center}

\begin{thebibliography}{99}
%
\bibitem{data} D.\ Buskulic et al., ALEPH collab., 
Phys. Lett. {\bf B374}, 319 (1996); paper submitted to the 
``XVIII International Symposium on Lepton Photon Interactions'', 
1997, Hamburg, Germany, paper no.\ {\bf LP279}; \\
DELPHI collab., DELPHI 95-86 PHYS 521 (paper submitted to
the EPS-HEP 95 conference, Brussels, 1995); \\
K.\ Ackerstaff et al., OPAL collab., CERN-PPE/97-104, 
{\tt hep-ex/9708027}. 
%
\bibitem{dsv} D. de Florian, M. Stratmann and W. Vogelsang, to be
published in Phys. Rev. {\bf D}, {\tt hep-ph/9711387}.
%
\bibitem{rhic} RHIC Spin Collab., D.\ Hill et al., letter of intent 
               RHIC-SPIN-LOI-1991, updated 1993;\\
               G. Bunce et al., Particle World {\bf 3}, 1 (1992); \\
               PHENIX/Spin Collaboration, K.~Imai et al., 
               BNL-PROPOSAL-R5-ADD (1994).
%
\bibitem{heran} V.A. Korotkov and W.-D. Nowak, DESY-97-004,
                {\tt hep-ph/9701371}, talk presented at the 2nd ELFE
                workshop, Saint\ Malo, France, Sept.\ 1996; \\
                M. Anselmino et al., DESY-Zeuthen 96-04 (1996);
                Proc. of the Workshop on ``Future Physics at HERA'', 
                Hamburg, 1995, p.837, {\tt hep-ph/9608393}.  
%
\bibitem{gw} R. Gastmans and T.T. Wu, ``The Ubiquitous Photon''  
(Clarendon Press - Oxford, 1990).
%
\bibitem{singlepol} D.\ de Florian, C.\ Garc\'\i a Canal, and R.\ Sassot, 
Nucl. Phys. {\bf B470}, 195 (1996).
%
\bibitem{doublepol} D.\ de Florian and R.\ Sassot, 
Nucl. Phys. {\bf B488}, 367 (1997).
%
\bibitem{thesis} M.\ Stratmann, Ph.D.\ Thesis, Univ.\ Dortmund
report DO-TH 96/24 (unpublished).
%
\bibitem{rapidity} see, for example, M.\ Gl\"uck and W.\ Vogelsang,
Phys. Lett. {\bf B277}, 515 (1992).
%
\bibitem{burkjaf} M.\ Burkardt and R.L.\ Jaffe, 
Phys. Rev. Lett. {\bf{70}}, 2537 (1993).
%
\bibitem{jaffe2} R.L.\ Jaffe, Phys. Rev. {\bf{D54}}, 6581 (1996).
%
\bibitem{gej} M.\ Gourdin, Nucl. Phys. {\bf{B38}}, 418 (1972);\\
J.\ Ellis and R.L.\ Jaffe, Phys. Rev. {\bf{D9}}, 1444 (1974);
{\bf{D10}}, 1669 (E) (1974).
%
\bibitem{grv} M.\ Gl\"uck, E.\ Reya, and A.\ Vogt, 
Z. Phys. {\bf C67}, 433 (1995).
%
\bibitem{grsv} M.\ Gl\"uck, E.\ Reya, M.\ Stratmann, and W.\ Vogelsang, 
Phys. Rev. {\bf D53}, 4775 (1996).
%
\bibitem{dss} D. de Florian, O. Sampayo and R. Sassot, to be published in 
Phys. Rev. {\bf D}, {\tt hep-ph/9711440}.
%
\end{thebibliography}
\end{document}